\documentclass[%
 article,
 amsmath,amssymb,
 reprint,%
 longbibliography,
]{revtex4-1}

\usepackage{graphicx}% Include figure files
\usepackage{dcolumn}% Align table columns on decimal point
\usepackage{bm}% bold math
\usepackage[utf8]{inputenc}
\usepackage[T1]{fontenc}
\usepackage{mathptmx}
\usepackage{color}
\usepackage{amsmath}
\usepackage[ruled,lined]{algorithm2e}

\usepackage[page]{appendix}

\usepackage{caption}
\usepackage{subcaption}

\usepackage{graphicx}

\usepackage{tikz}
\usepackage{pgfbaselayers}
\usetikzlibrary{quantikz}

\usepackage[textwidth=7in,top=1in,bottom=1.1in,tmargin=0.4in]{geometry}
\usepackage{fancyhdr}
\usepackage{multirow}
\usepackage{booktabs}
\begin{document}

\pagestyle{fancy}
\fancyhf{}
%\rfoot{IonQ Proprietary Information}
\cfoot{\thepage}

%\title[]{Correlating Electronic Pairs with Variational Quantum Eigensolver on Trapped-Ion Quantum Computers}
\title[]{Orbital-optimized pair-correlated electron simulations on trapped-ion quantum computers}
\author{Luning Zhao}
\email{zhao@ionq.co}
\affiliation{
IonQ Inc, College Park, MD, 20740, USA
}

\author{Joshua Goings}
\affiliation{
IonQ Inc, College Park, MD, 20740, USA
}

\author{Kyujin Shin}
\email{shinkj@hyundai.com}
\affiliation{
 Materials Research \& Engineering Center, R\&D Division, Hyundai Motor Company, Uiwang 16082, Republic of Korea
}

\author{Woomin Kyoung}
\affiliation{ 
  Materials Research \& Engineering Center, R\&D Division, Hyundai Motor Company, Uiwang 16082, Republic of Korea
}
\author{Johanna I. Fuks}
\affiliation{Qunova Computing, Daejeon, 34051, Republic of Korea}
\author{June-Koo Kevin Rhee}
\affiliation{Qunova Computing, Daejeon, 34051, Republic of Korea}
\affiliation{School of Electrical Engineering, KAIST, Daejeon, 34141, Republic of Korea}
\author{Young Min Rhee}
\affiliation{Department of Chemistry, KAIST, Daejeon, 34141, Republic of Korea}
\author{Kenneth Wright}
\affiliation{
IonQ Inc, College Park, MD, 20740, USA
}
\author{Jason Nguyen}
\affiliation{
IonQ Inc, College Park, MD, 20740, USA
}
\author{Jungsang Kim}
\affiliation{
IonQ Inc, College Park, MD, 20740, USA
}

\author{Sonika Johri}
\affiliation{
IonQ Inc, College Park, MD, 20740, USA
}
\date{\today}

\begin{abstract}
%Solving the electronic structure problem has the potential to revolutionize a variety of industrial applications, from pharmaceutical drug discovery, to \textit{de-novo} battery design, to discovering new renewable energy sources, and even inventing new tools to mitigate climate change. Over the past several decades, myriad classical algorithms have been developed to solve the electronic structure problem. However, the most accurate methods have high polynomial, if not exponential, computational scaling, so reliably simulating systems with hundreds to thousands of atoms at high accuracy remains out of reach. Worse still, these classical algorithms are often ill-suited for treating strongly correlated systems, rendering detailed study of chemical reactivity, catalysis, and many other phenomena extremely challenging. In contrast, quantum algorithms for the solution of the electronic structure problems have emerged as a route to efficiently simulate chemistry at high accuracy. In particular, the variational quantum eigensolver (VQE) is one of the most promising algorithms for near-term quantum computers. 

Variational quantum eigensolvers (VQE) are among the most promising approaches for solving electronic structure problems on near-term quantum computers.
A critical challenge for VQE in practice is that one needs to strike a balance between the expressivity 
of the VQE ansatz versus the number of quantum gates required to implement the ansatz, given the reality of noisy quantum operations on near-term quantum computers. In this
work, we consider an orbital-optimized pair-correlated approximation to the unitary coupled cluster with 
singles and doubles (uCCSD) ansatz and report a highly efficient quantum circuit implementation for trapped-ion architectures. We show that orbital optimization can recover significant additional electron correlation energy without sacrificing efficiency through measurements of low-order
reduced density matrices (RDMs). In the dissociation of small molecules, the method gives qualitatively accurate predictions in the strongly-correlated regime
when running on noise-free quantum simulators. On IonQ's Harmony and Aria trapped-ion quantum
computers, we run end-to-end VQE algorithms with up to 12 qubits and 72 variational
parameters - the largest full VQE simulation with a correlated wave function on quantum hardware. We find that even without error mitigation techniques, the predicted relative energies
across different molecular geometries are in excellent agreement with noise-free simulators. 

\end{abstract}

\maketitle

\section{Introduction}
Finding accurate solutions to the electronic structure problem is of great importance to various industries, from modeling pharmaceutical drug docking \cite{Holzmann22_00551}, to designing new materials for light harvesting and CO$_2$ reduction \cite{Troyer21_033055}, to elucidating reaction mechanisms in novel
battery materials \cite{Garcia21_134115}. However, the classical computational resources needed to solve the electronic structure problem exactly scales exponentially 
with the size of systems, which limits routine or practical application to systems with less than 20 electrons. To make the problem tractable on classical computers, various 
approximate approaches have been developed, each with different trade-offs between cost and accuracy. These approaches include density 
functional theory (DFT)\cite{Yang89_book}, coupled cluster (CC)\cite{Szabo-Ostland} methods, density matrix renormalization group methods (DMRG)\cite{schollwock05_259}, and quantum Monte Carlo methods (QMC)\cite{Zhao20_174105}. These methods are routinely applied toward
computational chemistry calculations both in academia and in industry. 

Despite the abundance of different classical approximations, the electronic structure problem is far from being solved. For example, systems with
strongly correlated electronic structure are notoriously difficult to solve. These systems are commonly encountered during bond breaking and formation, as well as when studying systems such as transition-metal-containing catalysts, large $\pi$-conjugated systems, and high-temperature superconductors. In these cases, approximate approaches may either fail completely (such as in single-reference methods like DFT or CC), or will be prohibitively costly (such as in multi-reference methods like DMRG or QMC). It is possible that
approximate classical approaches will never reliably solve the strong correlation problem. 

In contrast, quantum computation\cite{Chuang00_book} has attracted significant attention for its potential to solve certain computational problems more efficiently than with classical computers, 
especially since IBM launched the first cloud accessible quantum computer and Google
demonstrated quantum advantage\cite{Martinis19_505}.
One of its most promising applications is to solve electronic structure problems efficiently\cite{Aspuru-Guzik19_10856}: to illustrate, consider that for a problem containing $N$ spin orbitals, the number of classical
bits required to represent the wave function scales combinatorially with $N$, while on a quantum computer only $N$ qubits are needed. The exponential advantage
offered by quantum computers has motivated a great deal of research in developing quantum algorithms to solve the electronic structure problem. 

Of these, the variational quantum eigensolver (VQE) algorithm \cite{Brien14_5213, Martinis16_031007, Gambetta17_23879,Google20_1084,Kim20_33} is designed specifically for current near-term intermediate scale quantum (NISQ) computers. VQE estimates the ground
state of a system by implementing a shallow parameterized circuit, which is classically optimized to variationally minimize the energy expectation value. %Unlike QPE, which uses a defined (fixed) system-specific time propagator, 
The VQE algorithm allows the user to select
the form of the parameterized circuit. This flexibility allows one to adjust the circuit depth based 
on the quantum gate fidelity, number of qubits, and desired accuracy. 
%Less expressible, but shallower circuits can be run on low-fidelity QPUs, while more expressible, deeper
%circuits can be run on high-fidelity QPUs. 
This makes VQE especially suitable for the NISQ era.  

There is, however, no free lunch and the ability to run shallower circuits within the VQE comes with two costs. First, the predicted energy in most cases remains approximate, because the accuracy depends on the expressivity of the circuit form. Second, one needs to perform a large number of measurements
for VQE. %compared to QPE. 
This makes the choice of the ansatz perhaps the most important building block in the VQE algorithm. So what does one choose? Early
demonstrations of VQE on quantum hardware utilized the physically-motivated unitary coupled cluster with singles and doubles (uCCSD) ansatz\cite{Mayhall20_1,Mayhall19_3007}. uCCSD is well-known in the quantum chemistry community to be able to treat strongly correlated systems, while remaining classically intractable. As such, A. Peruzzo \emph{et al.}~\cite{Brien14_5213} used the uCCSD ansatz in the first VQE demonstration on a photonic quantum computer to solve for
the H$_2$ molecule in a minimal basis. O'Malley and coworkers\cite{Martinis16_031007} performed the same simulations on a superconducting
quantum computer with two qubits. In 2019, McCaskey\cite{Pooser19_99} and co-workers simulated metal hydrides
in a 2-electron, 2-orbital active space using the uCCSD ansatz on IBM's superconducting
quantum computers with four qubits. However, going beyond a minimal active space poses 
difficulties due to the rapid increase in the number of entangling gates for the uCCSD ansatz. 
The number of entangling gates in uCCSD (e.g. $CNOT$) scales as $O(N^4)$, where $N$ is the number of qubits. Even the most efficient 
implementation of uCCSD circuits contain thousands of entangling gates for small systems\cite{Duncan20_10515}, which makes it impractical to run on
NISQ quantum computers. 

Due to the impracticality of the uCCSD ansatz on NISQ quantum computers, hardware efficient ansatzes (HEA)\cite{Gambetta17_23879,Taverneli18_022322,Izmaylov18_6317,Izmaylov20_1055,Parrish21_113010} have attracted significant attention. Compared to the uCCSD ansatz, HEAs need significantly shallower circuits. 
In 2017, researchers from IBM published the first study\cite{Gambetta17_23879} of using HEAs 
on superconducting 
quantum computers to simulate H$_2$, LiH, and BeH$_2$ with 2, 4, and 6 qubits. However, noise in the quantum processing unit (QPU) led to unphysical behavior in the
predicted dissociation curve. In 2019, the
same researchers\cite{Gambetta19_491} use HEAs to demonstrate quantum error mitigation using the zero-noise
extrapolation technique with 4 qubits. In 2021, researchers\cite{Yamamoto21_70} used HEAs to study thermally
activated delayed fluorescence (TADF) with 2 qubits on superconducting quantum computers. 
They found that without using an unscalable error mitigation approach, even a 2-qubit
circuit yields qualitatively inaccurate predictions to the relative energy. 

The largest VQE
simulation performed on quantum hardware so far is the Hartree-Fock (HF) study by
Google \cite{Google20_1084}, in which they used a superconducting quantum computer to simulate the HF wave
function for hydrogen chains up to 12 qubits and 72 entangling gates. However, the calculations
faced a considerable amount of hardware noise, necessitating the use of Hartree-Fock specific error mitigation techniques to achieve sufficiently accurate results. A more recent study by Google\cite{Rubin22_10799} simulated a cyclobutene ring on a superconducting quantum computer with up to 10 qubits using pair-correlated
wave functions. Here, the ansatz was classically pre-optimized on a simulator, leaving the final energy evaluation to be executed on the quantum device. Despite this, this calculation still required classical error mitigation techniques to achieve reasonable results for the final quantum energy evaluation.

%The accumulation of noises on quantum hardware prevents one from studying larger systems
%with more complex wave functions. 

Trapped-ion quantum computers have several unique advantages over other currently available quantum computing architectures. First, the gate fidelity for trapped-ion qubits is typically higher than for superconducting qubits, which enables users to run deeper circuits. Second, trapped-ion qubits
are all-to-all connected. This means one is able to entangle arbitrary pairs of qubits without
performing expensive $SWAP$ operations to entangle non-adjacent qubits, which is usually
required on systems with interactions that do not form a complete graph. Although both of these advantages should
lead to higher fidelity when running VQE circuits, implementations of VQE on trapped-ion
quantum computers are rare, and this is mainly due to comparatively limited availability of the trapped
ion quantum computing hardware versus superconducting quantum computers. 
In this work, we fill this gap by performing VQE simulations on two generations of 
trapped-ion quantum computers constructed by IonQ, Inc. 

We consider an approximate ansatz derived from the uCCSD ansatz: the unitary pair CCD (upCCD) ansatz, in which only 
paired double excitations are retained. This allows us to map the fermionic representation to electron pairs, known as the hard-core boson representation.
From this, we show that the unitary pCCD ansatz then requires half the number of qubits to encode the state vector as compared to the uCCSD ansatz. We then 
introduce the optimal circuit for implementing an arbitrary electron pair excitation in terms of the number of $CX$ gates. The energy
expression for the upCCD ansatz is derived, and we find that at most 3 circuits are needed to compute the energy expectation 
value, regardless of the size of the system. The shallow circuit structure, along with a constant low number of
measurements required, make the upCCD ansatz a perfect candidate on NISQ quantum computers. 

The accuracy of the upCCD ansatz depends on the choice of the underlying orbitals. Previous studies on similar wave functions
have found that it is necessary to optimize the orbitals along with the cluster amplitudes, especially for strongly correlated 
systems. In this work, we find that the orbital optimization effects can be incorporated through classical post-processing, and only requires the 
measurements of one- and two-body reduced density matrices (RDM) of the upCCD ansatz on the quantum device. In our experiments, we observe that 
failure to use orbital optimization results in highly non-physical energy predictions in the 
bond-dissociation regime, but physical behavior can be fully recovered by optimizing orbitals together with
parameters in the upCCD ansatz. Due to the symmetry of the upCCD ansatz, the energy measurements automatically
yield the required measurements for RDMs. This allows us to improve the expressivity of 
the ansatz, especially for strongly correlated systems without increasing the circuit depth on the quantum computer.

In Table \ref{tab:techniques}, we list a collection of techniques used in the study, with inventions in
this work marked in \textit{italic}.
Our result (see Table \ref{tab:previous_vqe}) is the largest full VQE demonstration on a QPU using
a correlated wave function without error mitigation.

%that does not require any unscalable or non-general error mitigation techniques. 

\begin{table*}
\caption{
  Techniques and their corresponding effects used in the paper. 
  \label{tab:techniques}
}
\begin{tabular}{ c c c}
\hline\hline
 Technique & Effect & Previous work\\
 \hline
 Electron Pair$\rightarrow$Bosons & Reduce number of qubits by a factor of 2 & \cite{Kim20_33}\\  
 Givens Rotation with Magic Gate & Most efficient Givens rotation implementation in terms of $CX$ gates & \cite{Williams04_032315} \\
 Hamiltonian Grouping & 3 circuits per energy measurements regardless of system size & \cite{Gogolin21_032605} \\
 \textit{Measurement of RDMs} & \textit{3 circuits for all 1- and 2-RDMs regardless of system size} & this work  \\
 \textit{Orbital Optimization with Newton-Raphson} & \textit{Increase circuit expressivity without increasing depth} & this work\\
 \hline\hline
\end{tabular}
\end{table*}

\begin{table*}
\caption{
  Comparison between this work and previous publicly-reported VQE chemistry simulations on QPUs. 
  \label{tab:previous_vqe}
}
\begin{tabular}{ c c c c c >{\centering\arraybackslash}p{3cm} >{\centering\arraybackslash}p{2cm} >{\centering\arraybackslash}p{2cm}}
\hline\hline
 \multirow{2}*{Year} & \multirow{2}*{Ansatz} & \multirow{2}*{System} & \multirow{2}*{\# Qubits} & \multirow{2}*{\# Parameters} & \multirow{2}*{Full VQE?} & {Error Mitigation?} & Hardware Vendor \\
 \hline
 2022 & oo-upCCD [this work] & Li$_2$O & 12 & 72 & Yes & No & IonQ\\
 2022 & upCCD~\cite{Rubin22_10799} & Cyclobutene Ring & 10 & 25 & No & Yes & Google\\
 2022 & uCCSD~\cite{Singh22_14834} & CH$_3^{\cdot}$ & 6 & 4 & No & Yes & Quantinuum \\
 \multirow{2}*{2022} &  \multirow{2}*{YXXX~\cite{MunozRamo22_e26975}} &  \multirow{2}*{oxazine derivatives} &  \multirow{2}*{4} &  \multirow{2}*{1} & Yes~(superconducting), No~(trapped-ion) &  \multirow{2}*{Yes} & IBM, Quantinuum\\
 2022 & uCCSD-PBC~\cite{MunozRamo22_033110} & crystalline iron model & 2 & 1 & Yes & Yes & IBM \\
 2022 & Entang. Forging~\cite{Ohnishi22_02414} & H$_3$S$^{+}$ & 6 & 8 & No & Yes & IBM\\
 2022 & Entang. Forging~\cite{Sheldon22_010309} & H$_2$O & 5 & 3 & Yes & Yes & IBM\\
2021 & HEA~\cite{Garcia21_134115} & LiH (dipole moment) & 4 & 16 & Yes & Yes & IBM\\
 2021 & HEA~\cite{Yamamoto21_70} & TADF & 2 & 4 & Yes & Yes & IBM \\
 2021 & qubit CC~\cite{Yamazaki21_245} & H$_{10}$ & 2 & 3 & No & Yes & IonQ\\
 2021 & HEA~\cite{Yamamoto21_1827} & Li$_2$O$_4$ model & 2 & 4 & Yes & Yes & IBM\\
 2020 & Hartree-Fock~\cite{Google20_1084} & H$_{12}$ & 12 & 36 & Yes & Yes & Google\\
 2020 & upCCD~\cite{Kim20_33} & H$_2$O & 3 & 3 & Yes & No & IonQ\\
 2019 & HEA~\cite{Gambetta19_491} & LiH & 4 & 20 & Yes & Yes & IBM\\
 2019 & reduced uCC~\cite{Pooser19_1} & NaH, KH, RbH  & 4 & 3 & No & Yes & IBM, Rigetti\\
 2018 & uCCSD~\cite{Siddiqi18_011021} & H$_2$ (excited states) & 2 & 1 & Yes & No & UCB, LBNL \\
 2017 & HEA~\cite{Gambetta17_23879} & BeH$_2$ & 6 & 30 & Yes & No & IBM \\
 2016 & uCCSD~\cite{Martinis16_031007} & H$_2$ & 2 & 1 & No (scan) & No & UCSB \\
 2014 & uCCSD~\cite{Obrien14_1} & HeH$^{+}$ & 2 & 6 & Yes & No & Univ. of Bristol \\
 \hline\hline
\end{tabular}
\end{table*}

The paper is structured as follows. We begin by introducing the upCCD ansatz, then discuss the mapping from electron pairs to Pauli matrices, 
and an efficient quantum circuit implementation of the ansatz. We then derive the energy expression for the upCCD ansatz. 
Having laid out the general formalism, we then introduce the orbital optimization of upCCD using RDMs and the 
Newton-Raphson algorithm. Results are presented on quantum simulators and IonQ's Harmony and Aria quantum computers for potential energy surface predictions of LiH, H$_2$O, and 
Li$_2$O molecule. All the VQE experiments on simulator and quantum computers are end-to-end VQE runs, 
which means we perform both parameter optimizations and final energy evaluations. We conclude with 
a summary of our findings and comments on future directions.

Readers are strongly encouraged to read “Methods” section
before “Results” section. In “Methods” section, we describe the details of the quantum computer hardware used to run VQE and the specifics of the molecular models used to generate the quantum simulation circuits. We heavily use the notations defined in “Methods” section throughout “Results” section.

\section{Results}

\subsection{The VQE Algorithm and Circuit}
The unitary pair Coupled Cluster double (upCCD) ansatz is 
\begin{equation}
    \label{eqn:pccd_ansatz}
    \left|\Psi_{\mathrm{u}\mathrm{pCCD}}\right>=e^{T-T^\dagger}\left|\mathrm{HF}\right>
\end{equation}
in which $T$ is the pair-double cluster operator, defined as
\begin{equation}
    T=\sum_{ia}{t_i^a a_{a\alpha}^\dagger a_{a\beta}^\dagger a_{i\beta}a_{i\alpha}}
\end{equation}
in which $i$ and $a$ are indices for occupied and unoccupied orbitals in the HF state. 
$a^\dagger_{p\alpha}$ ($a^\dagger_{p\beta}$) and $a_{p\alpha}$ ($a_{p\beta}$) are the 
Fermionic creation and annihilation operators in the $p$th spin up (down) orbital. 

Each exponentiation of the pair-excitation operator can be efficiently implemented with 
the following circuit, 
\begin{center}
\begin{quantikz}
& \gate{S} & \qw      & \targ{}  & \gate{Ry(\theta)} & \targ{}  & \gate{S^\dagger} & \qw \\
& \gate{S} & \gate{H} & \ctrl{-1} & \gate{Ry(\theta)} & \ctrl{-1} & \gate{H}         & \gate{S^\dagger} \\
\end{quantikz}
\end{center}
Once the circuit is defined, one needs to measure the energy expectation value $\left<\Psi_{\mathrm{upCCD}}|H|\Psi_{\mathrm{upCCD}}\right>$ for the second-quantized Hamiltonian 
$H$. Originally, there are $O(N^4)$ terms in $H$, in which $N$ is the number of qubits. However, a majority 
of them do not contribute to the energy since they break pair symmetry. After eliminating these 
terms, one finds that only 3 measurements are needed in the $X$, $Y$, and $Z$ basis respectively to compute
the energy, regardless of the system size. 

The upCCD ansatz defined in Equation \ref{eqn:pccd_ansatz} is not invariant to the choice of underlying orbitals. 
Previous studies \cite{Bultinck14_853,Scuseria15_214116,Neuscamman16_5841,Tavernelli20_124107} on similar wave functions have found that it is necessary to optimize the orbitals along with the cluster
amplitudes, especially for strongly correlated systems. The orbital optimized upCCD ansatz is
\begin{equation}
    \label{eqn:oo-upccd}
    \left|\Psi_{\mathrm{oo}-\mathrm{upCCD}}\right>=e^Ke^{T-T^\dagger}\left|\mathrm{HF}\right>
\end{equation}
in which there are two different sets of parameters: 1) circuit parameters in the cluster operator $T$; 2) orbital rotation parameters in the the orbital rotation operator $K$, which is defined as 
\begin{equation}
    \label{eqn:k_op}
    K=\sum_{p>q}{\sum_{\sigma}{K_{pq}(a^\dagger_{p\sigma}a_{q\sigma}-a^\dagger_{q\sigma}a_{p\sigma})}}
\end{equation}
where $K$ is an anti-Hermitian matrix and $\sigma$ indexes the spin. 

As shown in the Methods section, we find the optimal set of orbital rotation parameters $K_{pq}$ with 
the Newton-Raphson (NR) algorithm, in which the energy gradient and Hessian are measured on the quantum 
computer. Then, the effect of orbital rotation operators can be fully absorbed into
one- and two-electron integrals through integral transformation, which is done efficiently on classical 
computers. In this way doing orbital optimization does not increase the circuit depth or the number of measurements.

%\section{Quantum Subspace Expansion}
%\begin{equation}
%    HC=SCE
%\end{equation}

%\begin{equation}
%    \begin{split}
%        H_{ij}=\left<\Psi_0|O_i^{\dagger}HO_j|\Psi_0\right> \\
%        S_{ij}=\left<\Psi_0|O_i^{\dagger}O_j|\Psi_0\right> \\
%    \end{split}
%\end{equation}

%\begin{equation}
%    O_i=a^\dagger_pa_q, a^\dagger_pa_qa^\dagger_ra_s
%\end{equation}

%\begin{equation}
%    ^6\Gamma_{xprutk}^{yqsvwl}=\left<\Psi_0|a^\dagger_xa_ya^\dagger_pa_qa^\dagger_ra_sa^\dagger_ua_va^\dagger_ta_wa^\dagger_ka_l|\Psi_0\right>
%\end{equation}

%\begin{equation}
%    \begin{split}
%        n_p=a^\dagger_pa_p \\
%        1-n_p=a_pa^\dagger_p \\
%    \end{split}
%\end{equation}

%\section{Results}

\subsection{Experimental Example}
All the calculations and experiments were performed using IonQ's in-house quantum chemistry library, which facilitates the preparation and execution of quantum variational algorithms on IonQ's cloud simulators and QPUs. We used the PySCF~\cite{Chan18_e1340} software suite to compute the molecular integrals necessary to define the second-quantized Hamiltonian, as well as compute the classical FCI reference energies.
\subsubsection{LiH}
\begin{figure}[h!]
\centering
\includegraphics[width=8.5cm,angle=0,scale=1.0]{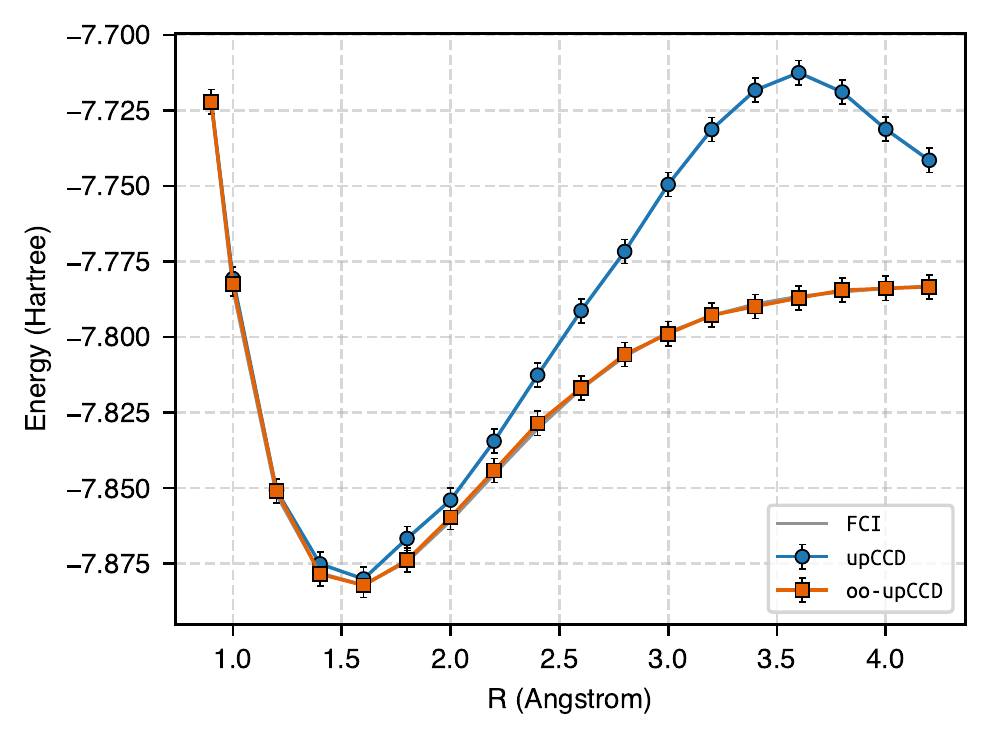}
\caption{
  Dissociation of LiH in STO-3G basis set comparing upCCD, oo-upCCD, and FCI. VQE results are obtained from a noise-free
  quantum simulator. 
        }
\label{fig:lih_simulator}
\end{figure}

We begin with our bond dissociation results with a simple example, the LiH molecule. The system has only two valence 
electrons. We freeze the Li 1s orbital, and also exclude the molecular orbitals formed with Li's 2px and 2py orbitals since
they do not contribute to the correlation energy due to symmetry. By doing so we only need 3 qubits and the VQE circuits 
consists of only 4 $CX$ gates. In Figure \ref{fig:lih_simulator}, we compare the energy predicted by FCI, upCCD, and oo-upCCD. 
As shown in the plot, both upCCD and oo-upCCD result in accurate energy predictions when the molecule is in the 
equilibrium geometry. However, as we enter the stretched region, upCCD gives highly nonphysical predictions. Not only does the 
energy error increases to tens of millihartrees, but it also exhibits a ``hump'' in the potential energy surface (PES). Such a 
nonphysical behavior is primarily due to the break down of the mean-field picture in stretched geometries. To the contrary, 
the oo-upCCD energy prediction matches FCI in both equilibrium and the stretched regions, which demonstrates the importance of
orbital optimization. 

\begin{figure}[h!]
\centering
\includegraphics[width=8.5cm,angle=0,scale=1.0]{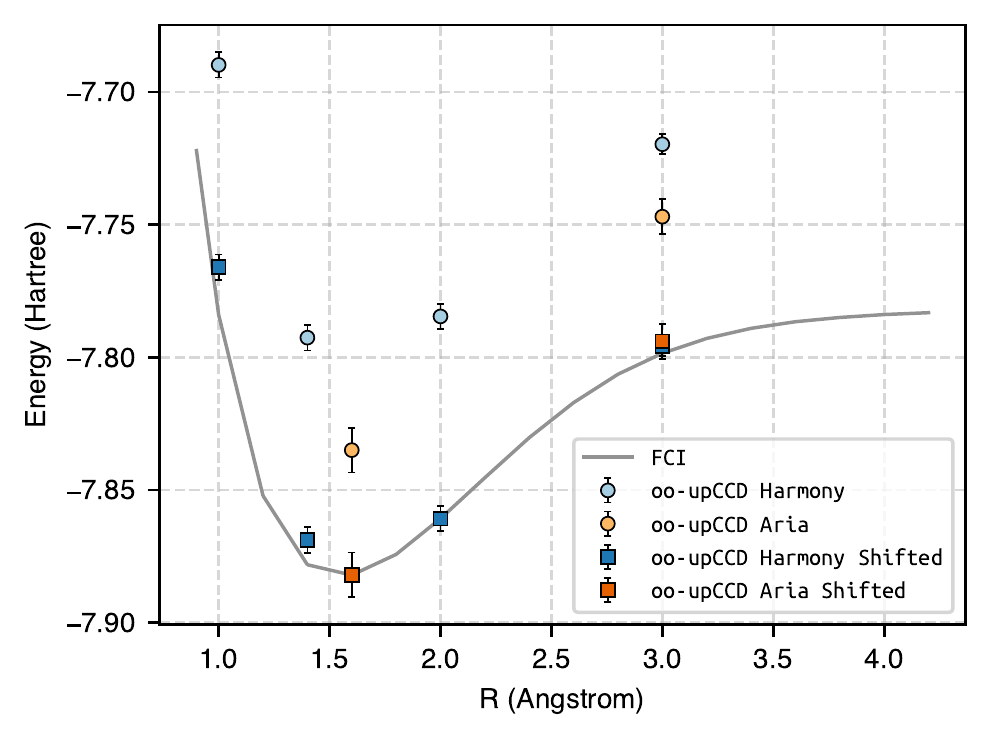}
\caption{
  Dissociation of LiH in STO-3G basis set comparing upCCD, oo-upCCD, and FCI. VQE results are obtained from the IonQ Harmony and the IonQ Aria quantum computer. 
        }
\label{fig:lih_aria}
\end{figure}

We then move from simulators to quantum hardware. In Figure \ref{fig:lih_aria}, we show the results obtained from IonQ's
Harmony quantum computer. The system has 11 all-to-all connected qubits, and the averaged single and two-qubit gate
fidelities are 99\% and 98\%. It has been used in numerous applications, including quantum chemistry \cite{Kim20_33,Yamazaki21_245}, quantum machine learning \cite{Kerenidis21_122,Ortiz22_031010}, 
and finance \cite{Wright21_06315,Zeng22_033034}. Due to the limited machine time, instead of scanning the entire PES, we selected a few points from squeezed, equilibrium, 
and stretched geometries. As shown in the plot, the energy measured on noisy quantum hardware is much higher than the 
simulation results. However, we also find that the amount of error is consistent along the PES. Based on such an observation, we 
shifted all the measured energies so that the energy at $R=1.6$ Angstrom matches the simulation energy. By doing so, 
the shifted energies matches well with the exact energy. This is notable especially with the stretched geometry $R=3.0$ Angstrom, in which the 
shifted energy falls on the simulated PES of oo-upCCD, demonstrating that the orbital optimization effects are successfully captured by the quantum hardware. 

%\begin{figure}[h!]
%\centering
%\includegraphics[width=8.5cm,angle=0,scale=1.0]{lih_harmony.pdf}
%\caption{
%  Dissociation of LiH in STO-3G basis set comparing upCCD, oo-upCCD, and FCI. VQE results are obtained from the IonQ Harmony quantum computer. 
%        }
%\label{fig:lih_s3/4}
%\end{figure}

Lastly, we ran the same simulation on IonQ Aria: IonQ's latest generation quantum computer and
the results are shown in Figure \ref{fig:lih_aria}. IonQ Aria offers both more qubits and improved gate 
fidelities over IonQ Harmony. We find that the improved gate fidelities 
reduces the amount of error in energy by 38\%. Once shifted, the relative 
energy also matches the exact energy within statistical uncertainty. The improvements
from Harmony to Aria are not very large in this case due to the simplicity of the 
circuit, which contains only 4 $CX$ gates. 

\subsubsection{H$_2$O}

\begin{figure}[h!]
\centering
\includegraphics[width=8.5cm,angle=0,scale=1.0]{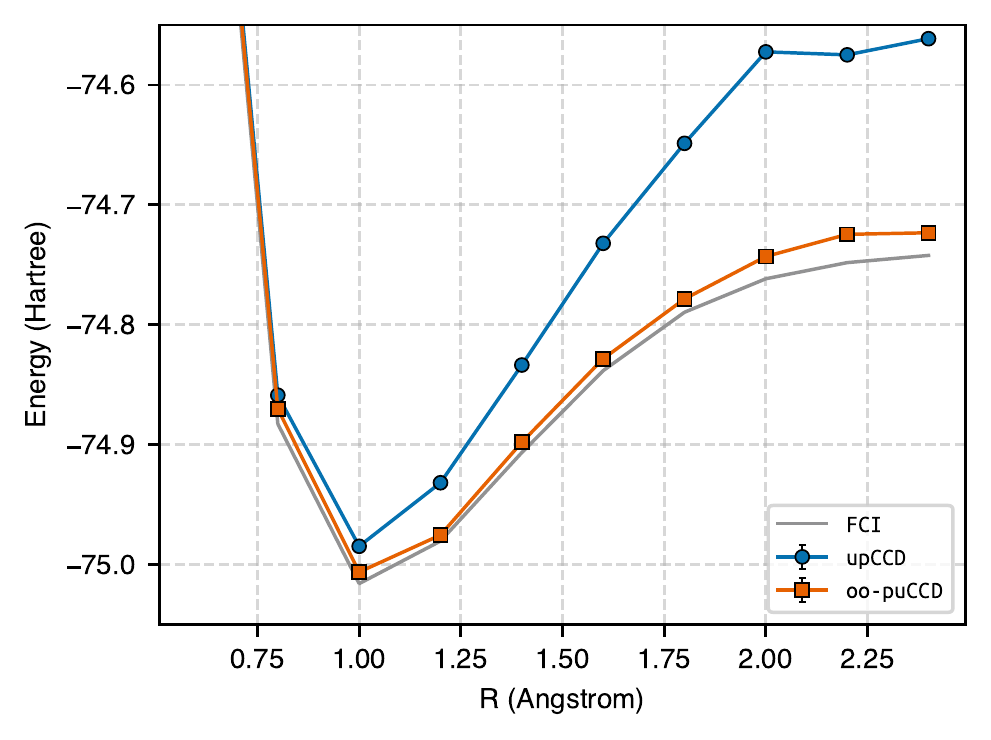}
\caption{
  Dissociation of H$_2$O in STO-3G basis set comparing upCCD, oo-upCCD, and FCI. VQE results are obtained from a noise-free
  quantum simulator. The statistical error from the 6000 shots is not visible at this scale. 
        }
\label{fig:h2o_simulator}
\end{figure}

Our next example is the symmetric double dissociation of H$_2$O, as shown in Figure \ref{fig:h2o_simulator} for results obtained 
on the simulator. We only freeze the O 1s core orbital and keep all other orbitals in the active space. The total number of qubits
required is 6 and there are 16 $CX$ gates in the circuit. 
Again, oo-upCCD produces highly accurate energies compared with FCI. However, unlike LiH, in which oo-upCCD matches
FCI exactly, in H$_2$O we find the predicted energy error for oo-upCCD is about 20 millihartrees, especially when we are in the 
stretched geometry. The error is due to the omission of the un-paired excitations in the oo-upCCD ansatz. However, we also note that
without orbital optimization, the upCCD ansatz using the HF orbitals yields more than 200 millihartrees of error in energy, again emphasizing 
the importance of orbital optimization. 

Before running the circuit on quantum hardware, we first remove redundant parameters from the
ansatz. The redundant parameters are the circuit parameters that do not contribute to the energy,
and their amplitudes stay zero during the optimization process. For the H$_2$O molecule, an example
of redundant parameters are the amplitudes that correspond to pair excitations from the non-bonding
orbital. In this study, we identify redundant parameters by tracking the evolutions of parameter
amplitudes on a noise-free simulator, with all parameters started from zero. Parameters whose 
amplitudes stay at zero during the entire optimization process are identified as redundant 
parameters. It is worth noting that such an approach does not scale as the system size, and the running time on simulator becomes prohibitively expensive.
Fortunately, there exist scalable approaches for identifying and simulating only non-redundant
parameters, such as the gradient based selection used for the ADAPT-VQE\cite{Mayhall19_3007} method.

\begin{figure}[h!]
\centering
\includegraphics[width=8.5cm,angle=0,scale=1.0]{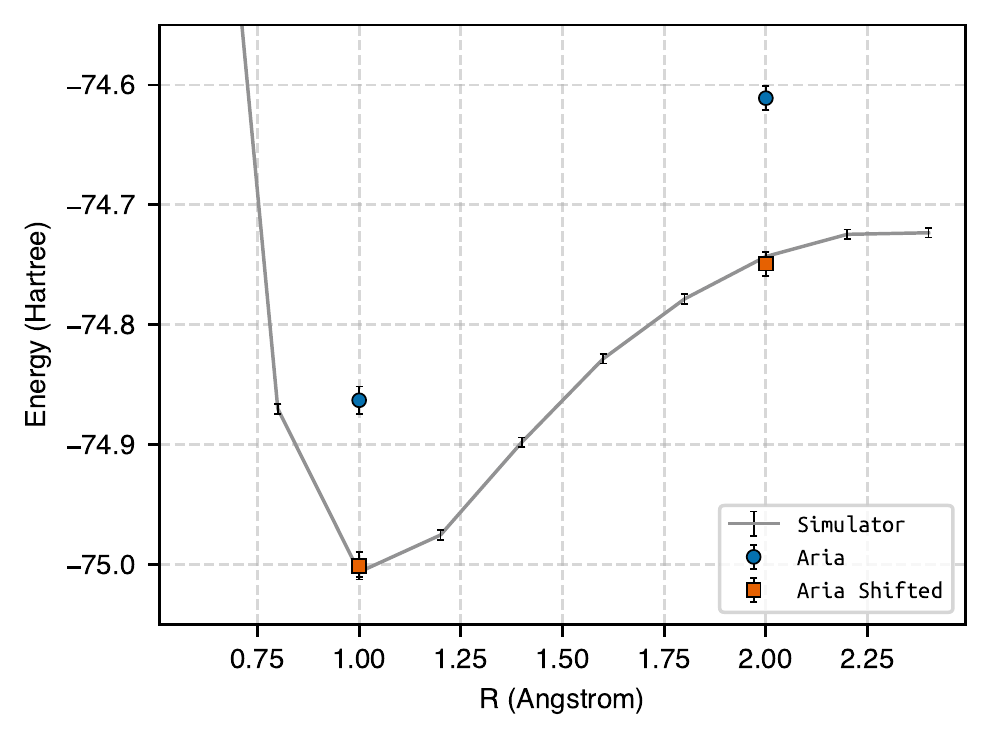}
\caption{
  Dissociation of H$_2$O in STO-3G basis set comparing oo-upCCD VQE results obtained from the IonQ Aria quantum computer and a noise-free simulator. 
        }
\label{fig:h2o_aria}
\end{figure}

Upon removal of redundant parameters, we are able to reduce the circuit to contain 4 circuit parameters
and 8 $CX$ gates. We then performed the oo-upCCD simulation on IonQ's Aria quantum computer, and the results are shown in Figure \ref{fig:h2o_aria}. The simulation is done on two geometry points: one at the equilibrium geometry and the other one at the stretched geometry. We find that
the Aria quantum computer successfully finds the optimal parameters and captures the orbital
optimization effects. Similar to LiH, the noise on the hardware introduce a systematic, 
positive bias to the measured absolute energy, but such a bias is constant at different geometry
points. Once we shift the absolute energies by a constant, the energies match the ones measured 
on a noise-free simulator, which demonstrates that the hardware noise is consistent enough so that the predicted relative energies are accurate. 

\subsubsection{Li$_2$O}

\begin{figure}[h!]
\centering
\includegraphics[width=8.5cm,angle=0,scale=1.0]{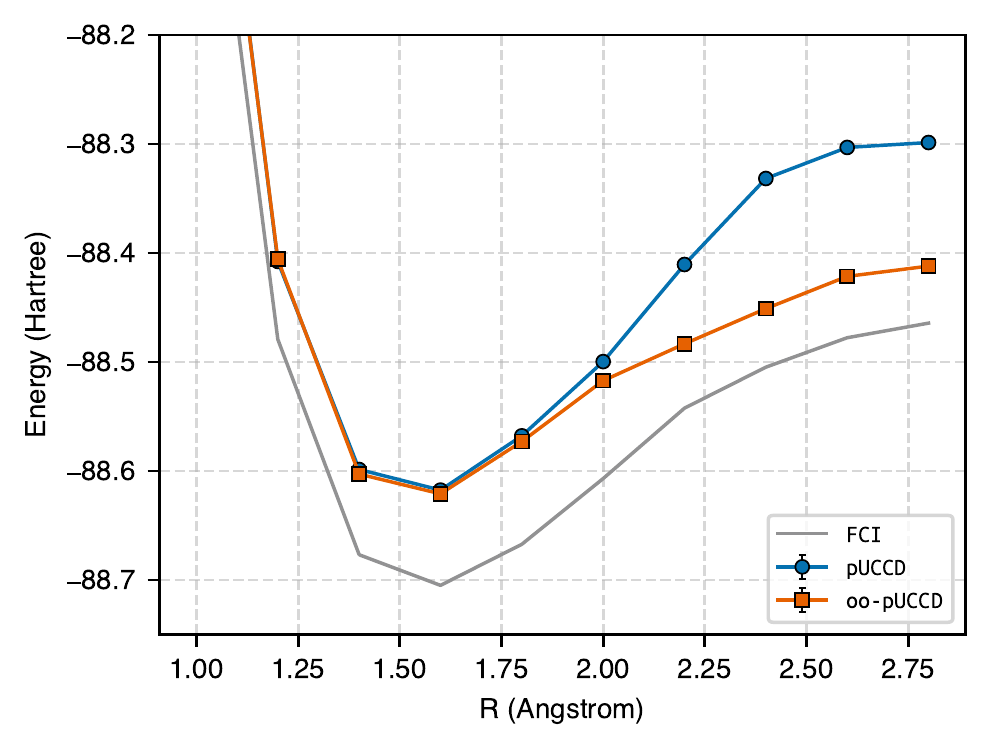}
\caption{
  Dissociation of Li$_2$O in STO-3G basis set comparing upCCD, oo-upCCD, and FCI. VQE results are obtained from a noise-free
  quantum simulator. The statistical error from the 6000 shots is not visible at this scale. 
        }
\label{fig:li2o_simulator}
\end{figure}

Our final example is the symmetric dissociation of the Li$_2$O molecule. Li$_2$O is one of the secondary reaction products in lithium-air batteries, 
which is believed to be a candidate for next-generation lithium battery due to its high energy density. We freeze the 1s orbital for 
Li and O, resulting in a circuit with 12 qubits and 64 $CX$ gates. The results on an ideal simulator are shown in Figure \ref{fig:li2o_simulator}. 
The difference in energy between oo-upCCD and FCI becomes more noticeable, which is expected since Li$_2$O has twice 
(four times) the number of orbitals as H$_2$O (LiH). As a result, there are many more unpaired excited configurations in Li$_2$O than
H$_2$O and LiH, and ignoring them, as is being done in the upCCD ansatz, introduces a more drastic approximation. Again, we find that 
orbital optimization does not make any noticeable amount of difference in equilibrium geometry, but becomes crucial in stretched 
geometries. 

\begin{figure}[h!]
\centering
\includegraphics[width=8.5cm,angle=0,scale=1.0]{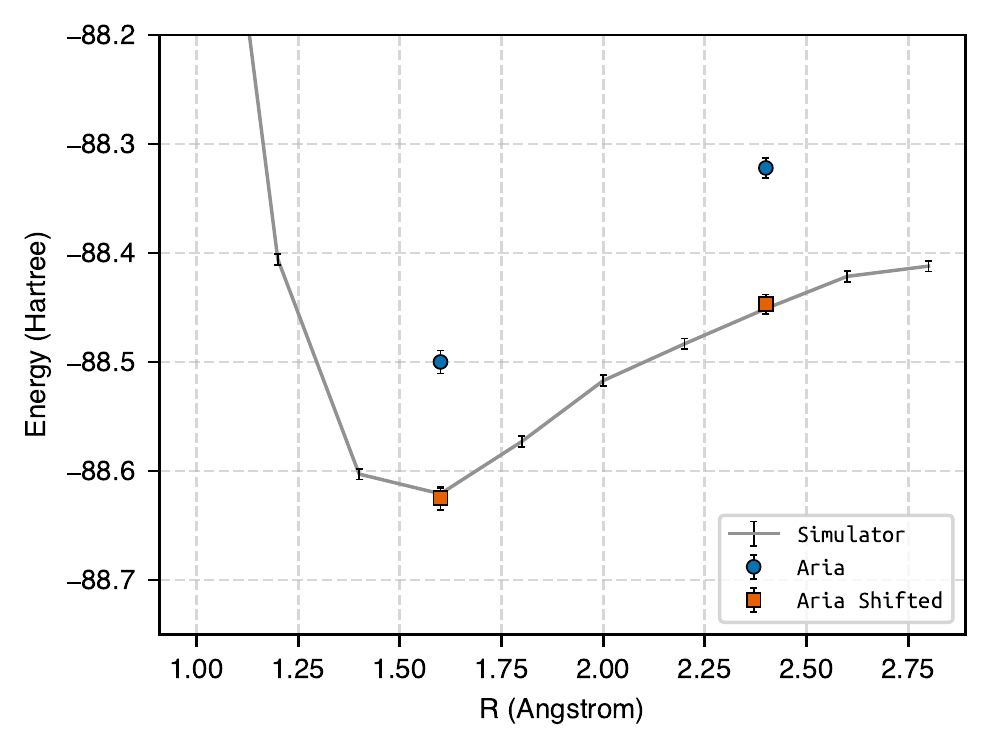}
\caption{
  Dissociation of Li$_2$O in STO-3G basis set comparing oo-upCCD VQE results obtained from the IonQ Aria quantum computer and a noise-free simulator. 
        }
\label{fig:li2o_aria}
\end{figure}

We then performed the oo-upCCD simulation on the Aria quantum computer. Analogous to H$_2$O, we first
identify and remove redundant parameters. In this example we find that only 6 out of the 32 circuit
parameters are non-redundant. Therefore, we only implement and optimize these 6 circuit parameters (12 $CX$ gates)
on the
quantum hardware, with an additional 66 orbital rotation parameters, for 72 variational parameters total. The results are shown in Figure \ref{fig:li2o_aria} at two geometry 
points: one at the equilibrium geometry and another one at the stretched geometry. Like in
H$_2$O, we find that despite hardware noise, the predicted relative energy matches the
simulator's prediction, and the orbital optimization effects are successfully captured by
Aria. 

\section{Discussion}
Quantum computers are expected to be able to efficiently solve the electronic structure problem. in principle, the electronic energy can be exactly computed in polynomial time using quantum phase
estimation (QPE)\cite{Head-Gordon05_1704} or its iterative variant\cite{white10_106}. In contrast, the best equivalent classical algorithm (full configuration interaction, or FCI) scales exponentially. In QPE, one implements the 
time propagator $U=\mathrm{exp}(-iHt)$ on the quantum computer and operates it
on an efficiently prepared trial state. Assuming the trial state has sufficient overlap with the 
exact eigenstate $\left|\Psi_i\right>$, the exact eigenstate's energy is encoded in the phase of the wave function since $U\left|\Psi_i\right>=\mathrm{exp}(-iE_it)\left|\Psi_i\right>$. The phase can be extracted using the quantum Fourier transform (QFT). 

While the QPE algorithm can compute the energy levels of molecules exactly, it is impractical on current NISQ computers. In the NISQ era, quantum gates are noisy, 
and entangling gates are typically an order of magnitude lower in fidelity compared to
single qubit gates. This means that one can only perform a limited number of quantum 
operations to ensure that the results are distinguishable from noise. This poses a significant
difficulty for the QPE algorithm, as the implementation of the time propagator is very
expensive and yields deep quantum circuits. QPE algorithms without using the time propagator, 
such as qubitization \cite{Neven18_041015,Chuang19_163,Babbush21_030305} have also been developed with improved scaling, but the fact remains that neither 
algorithm results in circuits that are shallow enough to run on quantum computers without error-corrected qubits.

We therefore focus on VQE, an algorithm expressly designed for NISQ computers. Here, we have developed an efficient VQE algorithm that is able to run on near-term quantum computers with high accuracy. The 
algorithm employs a chemically-inspired ansatz based on the unitary pair coupled cluster doubles (upCCD) wave function. The upCCD ansatz is obtained from
the general unitary CCSD ansatz by retaining only paired double excitations. This allows us to condense electron pairs to the hard-core boson representation
and develop an efficient quantum circuit implementation that only requires 2 $CX$ gates to implement one excitation. Since the 
accuracy of the upCCD ansatz depends on the underlying orbital choice, we developed an orbital optimization algorithm that 
finds the variationally optimal set of orbitals automatically. We find that orbital optimization can be implemented efficiently by measuring one- and two-body RDMs on a quantum computer and computing integral transformations on a classical computer.  

We tested the oo-upCCD VQE approach on the bond dissociation pathways for LiH, H$_2$O, and Li$_2$O molecule on both quantum simulators and
IonQ's Harmony and Aria quantum computers. We find that on quantum simulators,
oo-upCCD gives qualitatively accurate predictions to energy both in the 
weakly correlated and strongly correlated regime. However, upCCD without 
orbital optimization produces unphysical behavior in the strongly correlated
regime. On quantum hardware, we observed that noisy quantum gate
operations yield a consistent positive bias for the energy. Such a consistent bias has 
also been observed before \cite{Rubin22_10799}. In order to understand this, we have performed 
simulations with both coherent and incoherent noise models on quantum simulators. 
We find that if the error rate is low enough (below 1\%), both error models produce a constant additive error for different molecular geometries, which aligns with the error rate of the 
Harmony and Aria quantum computers. The simulation results can be found in the 
supplementary information. Therefore, although the measured absolute energies can be 
higher than simulator results by hundreds of millihartrees, the relative energies measured
are accurate due to the consistency of errors across the PES (i.e., low non-parallelity error).

As with other seniority zero approaches, oo-upCCD proves effective for describing some strong electron correlations but is unable to 
deliver quantitative accuracy, a difficulty that may in future be addressed in two different ways. First, one may consider implementing the full 
unitary-CCSD ansatz with quantum circuits, and pay the price of ending up
with very deep circuits that are not practical to run on NISQ devices, even
with the most efficient compilation techniques. A more practical way 
is to trade-off circuit depths with measurements and employ approaches 
such as the quantum subspace expansion (QSE) \cite{Siddiqi18_011021, McClean20_011004}. QSE will be able to solve 
two problems at the same time: 1) account for correlations contributed from 
non-bosonic excitations. and 2) account for correlations contributed from 
orbitals that are outside of the active space. QSE is able to achieve these
two goals without increasing circuit depth, by just performing more 
measurements to compute higher order RDMs. 

In order to achieve quantitative accuracy on a noisy quantum computer 
using VQE, one would inevitably perform some form of error mitigation. 
Over the past few years, various error mitigation methods have been developed, 
such as noise extrapolation \cite{Gambetta19_491}, density matrix purification \cite{Pooser19_99, Google20_1084}, symmetry
verification \cite{Benjamin19_180501}, randomized compiling \cite{Siddiqi21_041039}, and noise-estimation.\cite{Jong21_270502} We 
believe that an efficient VQE approach combined with measurement based 
post-processing and noise estimation is a very promising route that 
harvests the most performance out of near-term quantum computers. Together
with continued improvements in quantum hardware, both in terms of qubit
number and qubit quality, we will soon see quantum simulation of molecules
and materials that surpasses the best classical supercomputers. 
%These exciting research are underway. 

\section{Methods}
\subsection{Trapped-Ion Quantum Computer}

\begin{figure}[h!]
\centering
\includegraphics[width=8.5cm,angle=0,scale=1.0]{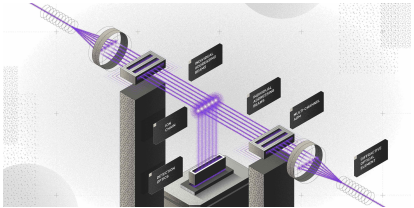}
\caption{Schematic of the IonQ Aria quantum computer, where two fans of individually
addressable beams illuminate a chain of individually imaged ions.}
\label{fig:qc}
\end{figure}

The experimental demonstration was performed on two generations of quantum processing units (QPU) from IonQ: Harmony and Aria. 
Both QPUs utilize trapped Ytterbium ions where two states in the ground
hyperfine manifold are used as qubit states. These states are manipulated by illuminating individual ions with
pulses of 355 nm light that drive Raman transitions between the ground states defining the qubit. By configuring these pulses, arbitrary single qubit gates and M{\o}lmer-S{\o}renson type two-qubit gates can both be realized. The IonQ Aria 
QPU (schematic in Figure 7) features not only an order of magnitude better performance in terms of fidelity but also is considerably more
robust compared to the IonQ Harmony QPU \cite{Proctor21_03137}.

\subsection{upCCD Circuit Design}
From the electron pair excitation operators, we can define the pair creation and annihilation operators
\begin{equation}
    \begin{split}
            &d_a^\dagger=a_{a\alpha}^\dagger a_{a\beta}^\dagger \\
            &d_i=a_{i\beta} a_{i\alpha} \\
    \end{split}
\end{equation}
in which $a_i$ and $a_i^\dagger$ are the fermionic annihilation and creation operators on orbital $i$. $\alpha$ and $\beta$ indicate 
spin up and spin down. 

They follow bosonic symmetries
\begin{equation}
    \begin{split}
        &[d_a^\dagger, d_i]=0 \\
        &[d_i,d_j]=[d_a^\dagger,d_b^\dagger]=0 \\
    \end{split}
\end{equation}

By performing the Jordan-Wigner Transformation (JWT) to map molecular orbitals to qubits, the pair creation and annihilation operators 
becomes
\begin{equation}
    \begin{split}
        d_a^\dagger\rightarrow \frac{1}{2}\left(X_a-iY_a\right) \\
        d_i\rightarrow \frac{1}{2}\left(X_i+iY_i\right) \\
    \end{split}
\end{equation}

The pair-excitation operator becomes 
\begin{equation}
    \label{eqn:pair_double_excitation_JWT}
    d_a^\dagger d_i\rightarrow\frac{1}{4}\left(X_aX_i+iX_aY_i-iY_aX_i+Y_aY_i\right)
\end{equation}

As one can see from the above equation, after JWT, the pair excitation operator does not have the Pauli-Z strings that occur 
in general double excitations, due to its bosonic nature.

The exponential of the pair-excitation operator, subtracted by 
its complex conjugate, becomes 
\begin{equation}
    \begin{split}
        &\mathrm{exp}\left(t_i^a(d_a^\dagger d_i-d_i^\dagger d_a)\right) \\
        &=\mathrm{exp}\left(\frac{it_i^a}{2}(X_aY_i-Y_aX_i)\right) \\
    \end{split}.
\end{equation}
One can then show that this is the Givens rotation matrix
\begin{equation}
    \begin{pmatrix}
            1 & 0 & 0 & 0\\
            0 & \mathrm{cos}(\frac{t_i^a}{2}) &  -\mathrm{sin}(\frac{t_i^a}{2}) & 0 \\
            0 & \mathrm{sin}(\frac{t_i^a}{2}) &  \mathrm{cos}(\frac{t_i^a}{2}) & 0 \\
            0 & 0 & 0 & 1 
\end{pmatrix}
\end{equation}

As has been shown before \cite{Williams04_032315}, the Givens rotation
matrix belongs to the \textbf{SO}(4) group, which can be implemented in 
12 elementary (i.e. $R_y$, $R_z$) gates and 2 $CX$ gates using 
the magic gate basis
\begin{center}
\begin{quantikz}
& \gate{S} & \qw      & \targ{} & \qw \\
& \gate{S} & \gate{H} & \ctrl{-1} & \qw \\
\end{quantikz}
\end{center}

The efficient Givens rotation implementation with angle $\theta$ is the following 
circuit. 
\begin{center}
\begin{quantikz}
& \gate{S} & \qw      & \targ{}  & \gate{Ry(\theta)} & \targ{}  & \gate{S^\dagger} & \qw \\
& \gate{S} & \gate{H} & \ctrl{-1} & \gate{Ry(\theta)} & \ctrl{-1} & \gate{H}         & \gate{S^\dagger} \\
\end{quantikz}
\end{center}
in which only two $CX$ gates are required. 

\subsection{Hamiltonian and Energy Measurements}
Since the upCCD ansatz conserves electron pairs, the terms in the $ab$ $initio$ Hamiltonian
that break electron pairs do not contribute to energy. After removing these terms, the 
Hamiltonian can be written as 
\begin{equation}
    H=H_1(n_p)+H_2(d^\dagger_pd_q)
\end{equation}
in which the first term only depends on the number operator
\begin{equation}
    n_p=a^\dagger_pa_p\rightarrow\frac{1}{2}(1-Z_p),
\end{equation}
and so it can be measured in the computational basis. 

The second term only depends on the pair excitation operator defined in Equation \ref{eqn:pair_double_excitation_JWT}. Furthermore, we note that the two middle terms in 
it are associated with purely imaginary coefficients, which do not contribute to energy,
so that this term can be measured with all qubits in either the $X$ or $Y$ basis. 
In summary, only 3 circuits are needed to be run in order to measure the energy expectation value for the upCCD ansatz, compared with a number of measurements that scales as $O(N^4)$ (where $N$ is the number of orbitals) if no symmetry is exploited, independent of the 
size of the system. This makes the upCCD ansatz extremely efficient in terms of number of measurements. 

\subsection{Orbital Optimization based on Measurements}
The orbital optimization effects can be performed classically with integral
transformation. Consider the energy expectation value of the oo-upCCD ansatz. 
\begin{equation}
    \label{eqn:eng_oo_upccd}
    E=\left<\Psi_{\mathrm{upCCD}}|e^{-K}He^K|\Psi_{\mathrm{upCCD}}\right>
\end{equation}

where $K$ is an anti-Hermitian operator defined in Equation \ref{eqn:k_op}. 
We first organize the elements of the lower triangle of $K$ into the length-$d$ vector $\vec{\kappa}$, 
where
\begin{equation}
    d=(n_o+n_v)(n_o+n_v-1)/2
\end{equation}
Starting from initial orbitals ($\vec{\kappa}=0$), we expand the energy out to second order in $\vec{\kappa}$ to obtain
\begin{equation}
    E(\vec{\kappa})\approx E(0)+\vec{\kappa}^T\vec{\omega}+\frac{1}{2}\vec{\kappa}^TQ\vec{\kappa}
\end{equation}
where the length-$d$ energy gradient $\vec{\omega}$ and the $d\times d$ energy Hessian $Q$ are given by
\begin{equation}
    \begin{split}
        \omega_x=\frac{\partial E(\vec{\kappa})}{\partial \kappa_x} \\
        Q_{xy}=\frac{\partial^2 E(\vec{\kappa})}{\partial \kappa_x \partial \kappa_y} \\
    \end{split}
\end{equation}
which in turn are functions of the spinless one- and two-electron reduced density matrices (RDM)
\begin{equation}
    \begin{split}
        &\gamma_{pq}=\left<\Psi|a_{p\alpha}^\dagger a_{q\alpha}+a_{p\beta}^\dagger a_{q\beta}|\Psi\right> \\
        &\Gamma_{pr}^{qs}=\left<\Psi|\frac{1}{2}a_{p\alpha}^\dagger a_{q\alpha}a_{r\alpha}^\dagger a_{s\alpha}+\frac{1}{2}a_{p\beta}^\dagger a_{q\beta}a_{r\beta}^\dagger a_{s\beta}\right. \\
        &\left.+a_{p\alpha}^\dagger a_{q\alpha}a_{r\beta}^\dagger a_{s\beta}|\Psi\right> \\
    \end{split}
\end{equation}

Since the spinless RDMs are in the form of expectation values, they can be measured on the quantum computer, and since we only need 1- and 2-RDMs, the cost for measuring them is 
the same as measuring the energy. 
Using $\vec{\omega}$ and $Q$, we can choose a $\vec{\kappa}$ that reduces the energy using the Newton-Raphson (NR) method, 
\begin{equation}
    \vec{\kappa}=-Q^{-1}\vec{\omega}
\end{equation}
At this point, if we continue with more NR steps until the energy is minimized, we need to implement $\mathrm{exp}(K)$ using quantum circuits, which 
increases the circuit depth. However, we can instead reset $\vec{\kappa}$ to zero by absorbing its effects into one- and two-electron 
integrals through standard molecular orbital transformation. At this point, another NR step can be taken, and the method can be iterated to 
convergence. In this way, since $\vec{\kappa}$ is always zero, we do not need to implement it with quantum circuits. The VQE algorithm is
shown in Algorithm \ref{alg:vqe}, and the detailed expressions for the orbital gradients 
and Hessians in terms of 1- and 2-RDMs, as well as an example of VQE convergence with respect to optimization iterations, can be found in the supplementary information. 

\begin{algorithm}
\caption{VQE Algorithm for oo-upCCD}\label{alg:vqe}
\While{Energy not converged}
{
 Optimize $T$ using SPSA\\
 Update circuit parameters based on the optimal $T^\ast$ \\
 Optimize $K$ using NR \\
 Update orbital parameters based on the optimal $K^\ast$ \\
 Rotate one- and two-electron integrals to the new orbital basis \\
 Set $K$ to be 0 \\
 Compute energy to check convergence \\
}
\end{algorithm}

%\subsection{Numerical Experimental Details}
%The upCCD and oo-upCCD circuits are implemented in the  Qiskit software platform,\cite{Qiskit} and all the simulator results are obtained from the qasm\_simulator
%provided by Qiskit. The hardware results are obtained from IonQ's Harmony and Aria quantum computers. The full configuration
%interaction results were obtained from Psi4.\cite{psi4} The O and Li 1s orbitals are frozen at the RHF level. We take 2000 shots for 
%LiH and H$_2$O and 6000 shots for Li$_2$O. The simultaneous perturbation stochastic approximation (SPSA) optimization 
%technique is used to optimized the VQE circuit parameters. All the calculations are performed in the STO-3G basis set provided
%by the PySCF\cite{pyscf} package. 

\section{Data Availability}
The data presented in this manuscript are available from the corresponding author upon reasonable request.

\begin{acknowledgments}
We thank the Hyundai Motor Company for funding this research through the Hyundai-IonQ Joint Quantum Computing Research Project. We thank Dr. Tae Won Lim and Dr. Seung Hyun Hong for enlightening discussions.  

\end{acknowledgments}

\clearpage
\bibliography{Journal_Short_Name, main}% Produces the bibliography via BibTeX.

\appendix

\end{document}

\pagestyle{fancy}
\fancyhf{}
%\rfoot{IonQ Proprietary Information}
\cfoot{\thepage}

\title[]{Supplementary Information: Orbital-optimized pair-correlated electron simulations on trapped-ion quantum computers}
\author{Luning Zhao}
\email{zhao@ionq.co}
\affiliation{
IonQ Inc, College Park, MD, 20740, USA
}

\author{Joshua Goings}
\affiliation{
IonQ Inc, College Park, MD, 20740, USA
}

\author{Kyujin Shin}
\email{shinkj@hyundai.com}
\affiliation{
 Materials Research \& Engineering Center, R\&D Division, Hyundai Motor Company, Uiwang 16082, Republic of Korea
}

\author{Woomin Kyoung}
\affiliation{ 
  Materials Research \& Engineering Center, R\&D Division, Hyundai Motor Company, Uiwang 16082, Republic of Korea
}
\author{Johanna I. Fuks}
\affiliation{Qunova Computing, Daejeon, 34051, Republic of Korea}
\author{June-Koo Kevin Rhee}
\affiliation{Qunova Computing, Daejeon, 34051, Republic of Korea}
\affiliation{School of Electrical Engineering, KAIST, Daejeon, 34141, Republic of Korea}
\author{Young Min Rhee}
\affiliation{Department of Chemistry, KAIST, Daejeon, 34141, Republic of Korea}
\author{Kenneth Wright}
\affiliation{
IonQ Inc, College Park, MD, 20740, USA
}
\author{Jason Nguyen}
\affiliation{
IonQ Inc, College Park, MD, 20740, USA
}
\author{Jungsang Kim}
\affiliation{
IonQ Inc, College Park, MD, 20740, USA
}

\author{Sonika Johri}
\affiliation{
IonQ Inc, College Park, MD, 20740, USA
}

\date{\today}
\maketitle

%\clearpage
\onecolumngrid

\section{EXPERIMENTAL METHODS}
\subsection{Single-qubit operation}
Single-qubit gates are implemented via Rabi oscillations between $\left|0\right>$ and 
$\left|1\right>$, using an SK1
composite pulse sequence. The transition is driven by a two-photon Raman process from
two separate beams generated by a pulsed 355 nm laser, where one beam is tightly focused
onto the ion so it can be manipulated independently from the rest of the chain. We prepare
arbitrary single-qubit states on the Bloch sphere by controlling the phase and pulse area
delivered by the individual addressing beam. Conventional randomized benchmarking
and gate set tomography techniques are used to characterize our single-qubit gates, and
we observe fidelities 99\% for Harmony and 99.94\% for Aria with good repeatability. 

\subsection{Two-qubit operation}
Two-qubit entangling gates are implemented via the Mølmer-Sørenson interaction, where
an amplitude-modulated laser pulse, composed of non-copropagating beams, achieves full
spin-motion decoupling at the end of the gate. One of the laser beams uniformly
illuminates the ions, while the other individually addresses the two particular ions involved
in the gate. By varying the pulse area through the laser intensities, we control the geometric
phase $\theta$ of the interaction, defined in terms of the Pauli-X matrices $X_i$
on the $i$th ion as
\begin{equation}
    XX_{ij}(\theta)=e^{-i\theta X_iX_j}
\end{equation}
We calibrate the the maximally entangling $XX$ gate with $\theta=\pi/4$, and $CX$ gates 
are implemented as
\begin{center}
\begin{quantikz}
& \gate{R_y(\pi/2)}      & \gate[wires=2][1cm]{XX(\pi/4)} & \gate{Rx(-\pi/2)} & \gate{Ry(-\pi/2)} &\qw \\
& \qw       & \qw & \gate{Rx(-\pi/2)} & \qw & \qw  \\
\end{quantikz}
\end{center}
in which 1 $XX$ gates are needed. 

\subsection{IonQ Harmony and Aria}

\section{THEORY METHODS}

\subsection{Comparison of Two Implementations of the Givens Rotation}
In the main text, the Givens rotation is implemented as 
\begin{center}
\begin{quantikz}
& \gate{S} & \qw      & \targ{}  & \gate{Ry(\theta)} & \targ{}  & \gate{S^\dagger} & \qw \\
& \gate{S} & \gate{H} & \ctrl{-1} & \gate{Ry(\theta)} & \ctrl{-1} & \gate{H}         & \gate{S^\dagger} \\
\end{quantikz}
\end{center}

Besides using the magic gate basis, another way of implementing the Givens rotation is to use 
the $R_{xx}$ gates using the following circuit, 
\begin{center}
\begin{quantikz}
& \gate{S^\dagger} & \gate[wires=2][1cm]{XX(\theta)} & \gate{S} & \gate[wires=2][1cm]{XX(-\theta)} & \qw & \qw \\
& \qw & \qw & \gate{S^\dagger} & \qw & \gate{S} & \qw \\
\end{quantikz}
\end{center}
in which two $R_{xx}$ gates are used.

The above circuit is more efficient than the circuit using magic gate basis on trapped-ion
quantum computers since the $XX$-Ising interactions are native. However, this requires 
calibration of arbitrary angle $R_{xx}$ gates on the hardware level, as was done
in IonQ's previous study \cite{Kim20_33}. On IonQ's Harmony and Aria quantum computers, only
the $R_{xx}$ gate with a fixed angle ($\pi/4$) is calibrated, and any arbitrary angle
$R_{xx}$ gate will be compiled to 2 $CX$ gates first, then further transpiled to 4 fixed angle
$R_{xx}$ gates, which results in two times more entangling gates operations than the circuit
using magic gate basis. This makes the latter circuit more efficient.

\subsection{Expectation values of the Hamiltonian}
In order to measure the energy of the upCCD ansatz on the quantum computer, we first write the $ab$ $initio$ Hamiltonian as
\begin{equation}
    \begin{split}
         H&=\sum_{pq}{g_{pq}\left(a^\dagger_{p\alpha}a_{q\alpha}+a^\dagger_{p\beta}a_{q\beta}\right)} \\
         &+\sum_{pqrs}{(pq|rs)\left(\frac{1}{2}a^\dagger_{p\alpha}a_{q\alpha}a^\dagger_{r\alpha}a_{s\alpha}+\frac{1}{2}a^\dagger_{p\beta}a_{q\beta}a^\dagger_{r\beta}a_{s\beta}\right)} \\
         &+\sum_{pqrs}{(pq|rs)a^\dagger_{p\alpha}a_{q\alpha}a^\dagger_{r\beta}a_{s\beta}} \\
    \end{split}
\end{equation}
in which $(pq|rs)$ is the usual two-electron coulomb integrals in $(11|22)$ order, 
and $g_{pq}$ are modified one-electron integral
\begin{equation}
    g_{pq}=h_{pq}-\frac{1}{2}\sum_r{(pr|rq)}
\end{equation}

For the upCCD wave function, only a subset of the Hamiltonian terms will contribute 
to the energy. The first part involves only number operators
\begin{equation}
    \begin{split}
        H_1&=\sum_p{g_{pp}\left(n_{p\alpha}+n_{p\beta}\right)} \\
        &+\sum_{pq}{(pp|qq)\left(\frac{1}{2}n_{p\alpha}n_{q\alpha}+\frac{1}{2}n_{p\beta}n_{q\beta}+n_{p\alpha}n_{q\beta}\right)} \\
        &+\sum_{p\neq q}{(pq|qp)\left(\frac{1}{2}n_{p\alpha}(1-n_{q\alpha})+\frac{1}{2}n_{p\beta}(1-n_{q\beta})\right)} \\
    \end{split}
\end{equation}
and an additional term
\begin{equation}
    H_2=\sum_{p\neq q}{(pq|pq)a^\dagger_{p\alpha}a_{q\alpha}a^\dagger_{p\beta}a_{q\beta}}
\end{equation}

For the first part of the Hamiltonian, we need to compute the following expectation values
\begin{equation}
    \begin{split}
        z_p=\left<n_{p\alpha}+n_{p\beta}\right> \\
        \Gamma_{pq}=\left<n_{p\alpha}n_{q\alpha}+n_{p\beta}n_{q\beta}\right> \\
        \Delta_{pq}=\left<n_{p\alpha}n_{q\beta}\right>.
    \end{split}
\end{equation}
For the first one, one can measure the $p$th qubit in the $Z$ basis, and multiply the measured results by 2. For the second term, one can measure 
both the $p$th and the $q$th qubit in the $Z$ basis, and if both measurements
give 1, this term is 2, otherwise this term is 0. For the third term, we also 
measure the $p$th and the $q$th qubit, and if both measurements give 1, this term
is 1, and 0 otherwise. 

Then the simple part of the Hamiltonian becomes 
\begin{equation}
    \begin{split}
        \left<H_1\right>&=\sum_p{g_{pp}z_p}+\sum_{pq}{(pp|qq)\left(\frac{1}{2}\Gamma_{pq}+\Delta_{pq}\right)} \\
        &+\sum_{p\neq q}{\frac{1}{2}(pq|qp)\left(z_p-\Gamma_{pq}\right)} \\
    \end{split}
\end{equation}

For $H_2$, we first re-arange terms 
\begin{equation}
    \begin{split}
        H_2&=\sum_{p\neq q}{(pq|pq)a^\dagger_{p\alpha}a_{q\alpha}a^\dagger_{p\beta}a_{q\beta}} \\
        &=-\sum_{p\neq q}{(pq|pq)a^\dagger_{p\alpha}a^\dagger_{p\beta}a_{q\alpha}a_{q\beta}} \\
        &=\sum_{p\neq q}{(pq|pq)a^\dagger_{p\alpha}a^\dagger_{p\beta}a_{q\beta}a_{q\alpha}} \\
        &=\sum_{p\neq q}{(pq|pq)d^\dagger_{p}d_{q}} \\
    \end{split}
\end{equation}
which, upon JWT, becomes 
\begin{equation}
    \begin{split}
        H_2&=\frac{1}{4}\sum_{p\neq q}{(pq|pq)\left(X_pX_q+iX_pY_q-iY_pX_q+Y_pY_q\right)} \\
        &=\frac{1}{2}\sum_{p> q}{(pq|pq)\left(X_pX_q+Y_pY_q\right)} \\
    \end{split}
\end{equation}
which can be measured by measuring qubit $p$ and $q$ and $X$ or $Y$ basis. 

As one can see, only 3 circuits are needed to be run in order to measure the energy expectation value for the upCCD ansatz, and it does not depend on the 
size of the system. This makes the upCCD ansatz extremely efficient in terms of number of measurements. 

\subsection{Analytical Expressions for Orbital Gradient and Hessian}

For completeness, we include here expressions for the oo-upCCD orbital rotation gradient and hessian. These should provide 
everything needed for the Newton-Raphson algorithm we use for orbital optimization. 

The energy can be written as
\begin{align}
\label{eqn:en_orb_rot_app}
E(\vec{\kappa})=\langle\Psi| e^{-\hat{K}} H e^{\hat{K}} |\Psi\rangle
\end{align}
with 
\begin{align}
\label{eqn:orb_rot_app}
\hat{K}=\sum _{ p>q }^{  }{ \sum _{ \sigma  }^{  }{ { K  }_{ pq }\left( { a }_{ p\sigma  }^{ \dagger  }{ a }_{ q\sigma  }-{ a }_{ q\sigma  }^{ \dagger  }{ a }_{ p\sigma  } \right)  }  }
\end{align}
where we evaluate the orbital rotation unitary transformation as exp$\left( \hat{K}  \right) $ and transform the integrals into this new basis. 

The orbital gradient is
\begin{align}
\label{eqn:orb_gra}
\begin{split}
&{ \left( \frac { \partial E  }{ \partial { K  }_{ pq } }  \right)  }_{ \vec{\kappa} =0 } \\
&=\mathcal{P}_{pq}\sum_{\sigma}{\left<{\left[ H,{a}_{p\sigma}^{\dagger}{a}_{q\sigma}\right]}\right>} \\ 
&=\mathcal{P}_{pq}\left(g_{up}\gamma_{uq}-g_{qu}\gamma_{pu}\right) \\
&+\mathcal{P}_{pq}\left((uv|tp)\Gamma_{uv}^{tq}+(up|tv)\Gamma_{uq}^{tv}-(uv|qt)\Gamma_{uv}^{pt}-(qv|tu)\Gamma_{pv}^{tu}\right) \\
\end{split}
\end{align}
where $\mathcal{P}_{pq}$ is a permutation operator $\mathcal{P}_{pq}=1-\left( p\leftrightarrow q \right)$, and Einstein summation is used. 

Similarly, the Hessian is
\begin{align}
\label{eqn:hessian:app}
\begin{split}
&{ Q }_{ pq,rs }={ \left( \frac { { \partial  }^{ 2 }E  }{ \partial { K  }_{ pq }\partial { K  }_{ rs } }  \right)  }_{ \vec{\kappa} =0 } \\
&=\frac { 1 }{ 2 } \mathcal{ P }_{ pq }\mathcal{ P }_{ rs }\sum _{ \sigma ,\tau  }^{  }{ \left< { \left[ \left[ H,{ a }_{ p\sigma  }^{ \dagger  }{ a }_{ q\sigma  } \right] ,{ a }_{ r\tau  }^{ \dagger  }{ a }_{ s\tau  } \right]  } \right>  } \\
&+\frac { 1 }{ 2 } \mathcal{ P }_{ pq }\mathcal{ P }_{ rs }\sum _{ \sigma ,\tau  }^{  }{ \left< { \left[ \left[ H,{ a }_{ r\sigma  }^{ \dagger  }{ a }_{ s\sigma  } \right] ,{ a }_{ p\tau  }^{ \dagger  }{ a }_{ q\tau  } \right]  } \right>  } \\
\end{split}
\end{align}

We obtain
\begin{align}
\label{eqn:orb_hessian_app}
\begin{split}
Q_{pq,rs}&=-\mathcal{P}_{pq}\mathcal{P}_{rs}\left(g_{sp}\gamma_{qr}+g_{ps}\gamma_{rq}\right) \\
&+\frac{1}{2}\mathcal{P}_{pq}\mathcal{P}_{rs}\left(\delta_{qr}g_{tp}\gamma_{ts}+\delta_{qr}g_{st}\gamma_{pt}+\delta_{ps}g_{qt}\gamma_{rt}+\delta_{ps}g_{tr}\gamma_{tq}\right) \\
&-\mathcal{P}_{pq}\mathcal{P}_{rs}\left((ut|qr)\Gamma_{ut}^{ps}+(qu|tr)\Gamma_{pu}^{ts}+(ur|qt)\Gamma_{us}^{pt}+(qr|tu)\Gamma_{ps}^{tu}\right.\\
&\left.+(ut|sp)\Gamma_{ut}^{rq}+(up|st)\Gamma_{uq}^{rt}+(su|tp)\Gamma_{ru}^{tq}+(sp|tu)\Gamma_{rq}^{tu}\right) \\
&+\mathcal{P}_{pq}\mathcal{P}_{rs}\left((up|vr)\Gamma_{uq}^{vs}+(ur|vq)\Gamma_{us}^{vq}+(qv|su)\Gamma_{pv}^{ru}+(sv|qu)\Gamma_{rv}^{pu}\right) \\
&+\frac{1}{2}\mathcal{P}_{pq}\mathcal{P}_{rs}\left(\delta_{qr}(uv|tp)\Gamma_{uv}^{ts}+\delta_{qr}(up|tv)\Gamma_{us}^{tv}+\delta_{qr}(uv|st)\Gamma_{uv}^{pt}\right. \\
&\left.+\delta_{qr}(sv|tu)\Gamma_{pv}^{tu}+\delta_{ps}(uv|qt)\Gamma_{uv}^{rt}+\delta_{ps}(qv|tu)\Gamma_{rv}^{tu}\right. \\
&\left.+\delta_{ps}(uv|tr)\Gamma_{uv}^{tq}+\delta_{ps}(ur|tv)\Gamma_{uq}^{tv}\right)
\end{split}
\end{align}
in which $\delta_{pq}$ is 1 if $p=q$ and 0 otherwise, and Einstein summation is also used.

\subsection{Numerical Experimental Details}
The upCCD and oo-upCCD circuits are implemented in the  Qiskit software platform,\cite{Qiskit} and all the simulator results are obtained from the qasm\_simulator
provided by Qiskit. The hardware results are obtained from IonQ's Harmony and Aria quantum computers. The full configuration
interaction results were obtained from Psi4.\cite{psi4} The O and Li 1s orbitals are frozen at the RHF level. On Harmony, we take 8192 shots for LiH. On Aria, we take 2000 shots for 
LiH and H$_2$O and 6000 shots for Li$_2$O. The simultaneous perturbation stochastic approximation (SPSA) optimization 
technique is used to optimized the VQE circuit parameters. All the calculations are performed in the STO-3G basis set provided
by the PySCF\cite{pyscf} package. The H$_2$O bond angle is fixed at 109.57$^{\circ}$. Error bars on the potential energy surface plots are due to finite samplings.

\subsection{Convergence of VQE}
\begin{figure}[h!]
\centering
\includegraphics[width=8.5cm,angle=0,scale=1.0]{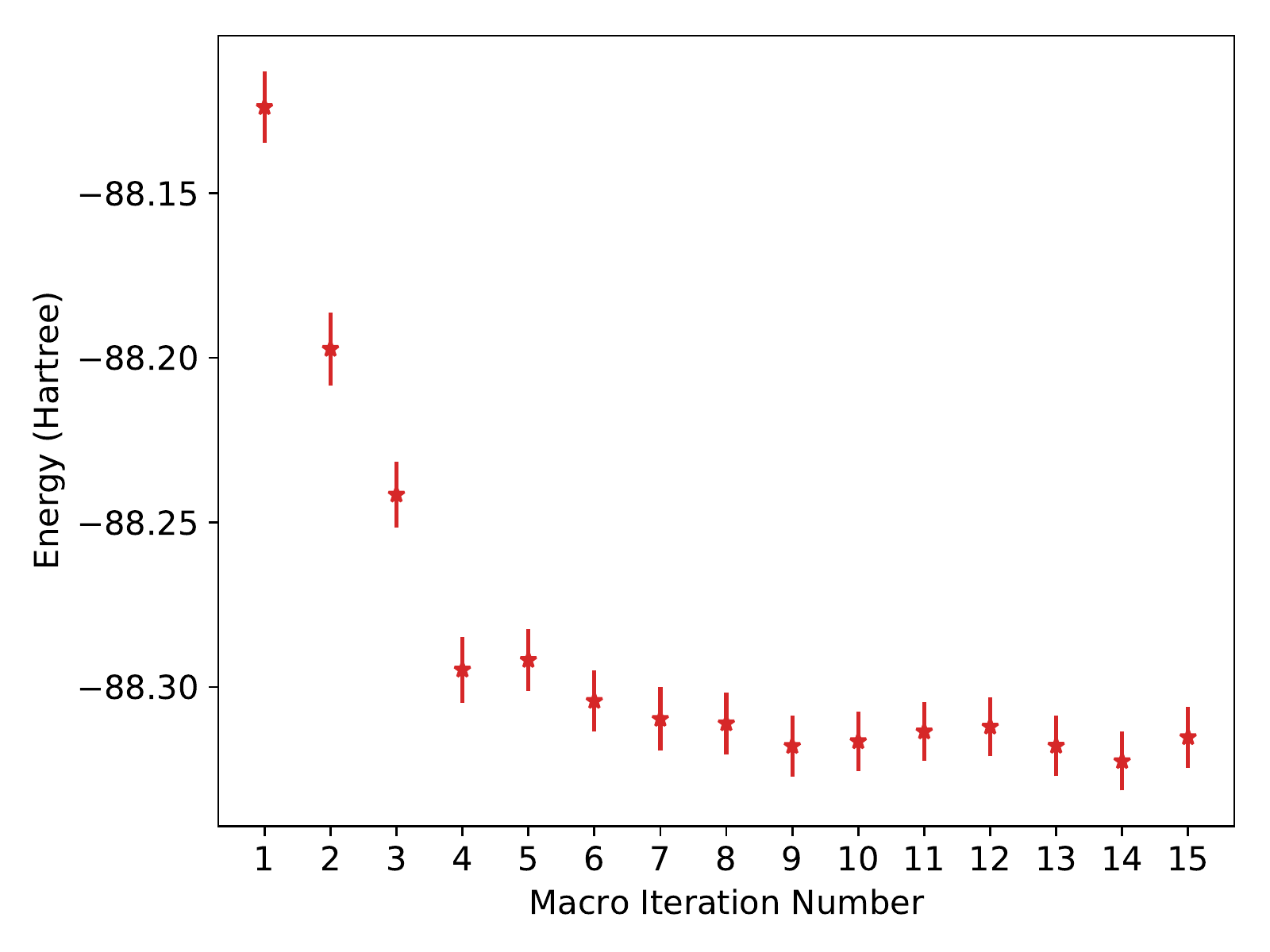}
\caption{
  Convergence of energy for the Li$_2$O molecule with $R=2.4$ Angstrom with respect to VQE iterations. 
        }
\label{fig:li2o_aria_convergence}
\end{figure}

The measured oo-upCCD energy on Aria for Li$_2$O as a function of the number of macro iterations that contain optimizations of both $T$ and $K$ is shown in Figure \ref{fig:li2o_aria_convergence}. As one 
could see, it converges in about 10 iterations. 

\subsection{Justification of Energy Shift}
In order to justify the constant energy shift we used in the main text, we performed numerical 
simulations for the upCCD circuit with different noise models. We consider the following two different
type of noise channelss. 

\subsubsection{Coherent Noise}
Coherent noise is simulated by changing the $CX$ gate implementation to 
\begin{center}
\begin{quantikz}
& \gate{R_y(\pi/2)} & \gate[wires=2][1cm]{ZX(r\delta)}     & \gate[wires=2][1cm]{XX(\pi/4+r\delta)} & \gate[wires=2][1cm]{ZX(-r\delta)} &  \gate{Rx(-\pi/2)} & \gate{Ry(-\pi/2)} &\qw \\
& \qw   & \qw  & \qw  & \qw & \gate{Rx(-\pi/2)} & \qw & \qw  \\
\end{quantikz}
\end{center}
in which $r$ is the error rate, and $\delta$ is a random variable sampled from $N(0,1)$. 

\begin{figure}[h!]
\centering
\includegraphics[width=8.5cm,angle=0,scale=1.0]{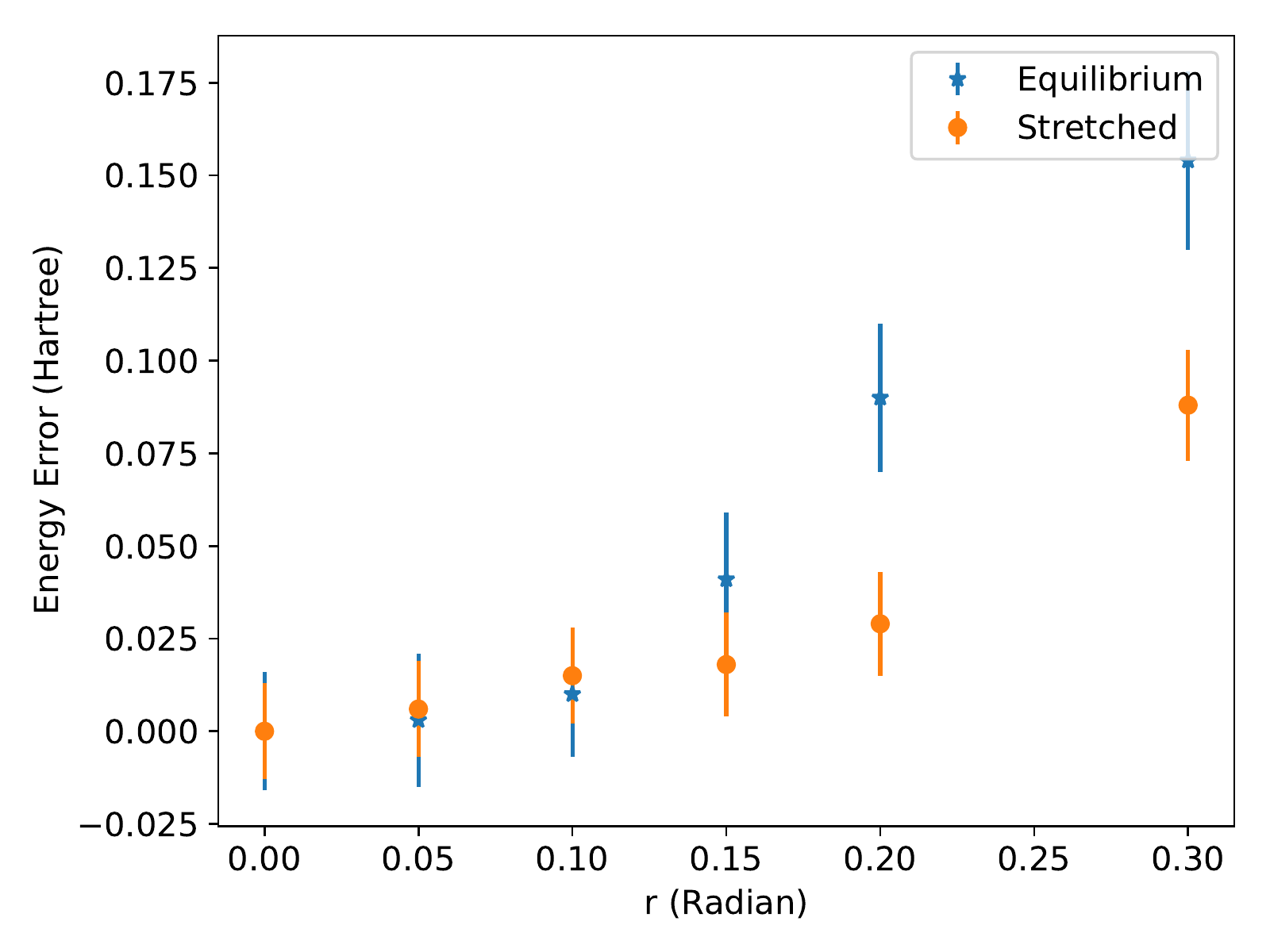}
\caption{
  VQE energy error as a function of $r$ in the coherent error model.
        }
\label{fig:coherent_error}
\end{figure}

We then use the above $CX$ implementation to simulate the $H_2$ molecule in the minimal basis set 
using the upCCD circuit with the $qasm$ simulator. We use 2000 shots and resample $\delta$ every
10 shots. Two different geometry points are used. The first one is $R=0.74$ Angstrom, in which $R$ 
is the distance between the two hydrogen atoms, and the second one is $R=1.74$ Angstrom. We will 
refer the first one as equilibrium and the second one as stretched. The simulate energy error v.s.
noise-free simulation is shown in Figure \ref{fig:coherent_error} . 

\subsubsection{Incoherent Noise}
Incoherent noise is simulated by using a depolarization noise channel
\begin{equation}
    \mathcal{E}(\rho)=(1-r)\rho+rI/2^N
\end{equation}
in which $\rho$ is the density matrix, $r$ is the error rate, $I$ is the identity matrix, and $N$ is 
the number of qubits.

\begin{figure}[h!]
\centering
\includegraphics[width=8.5cm,angle=0,scale=1.0]{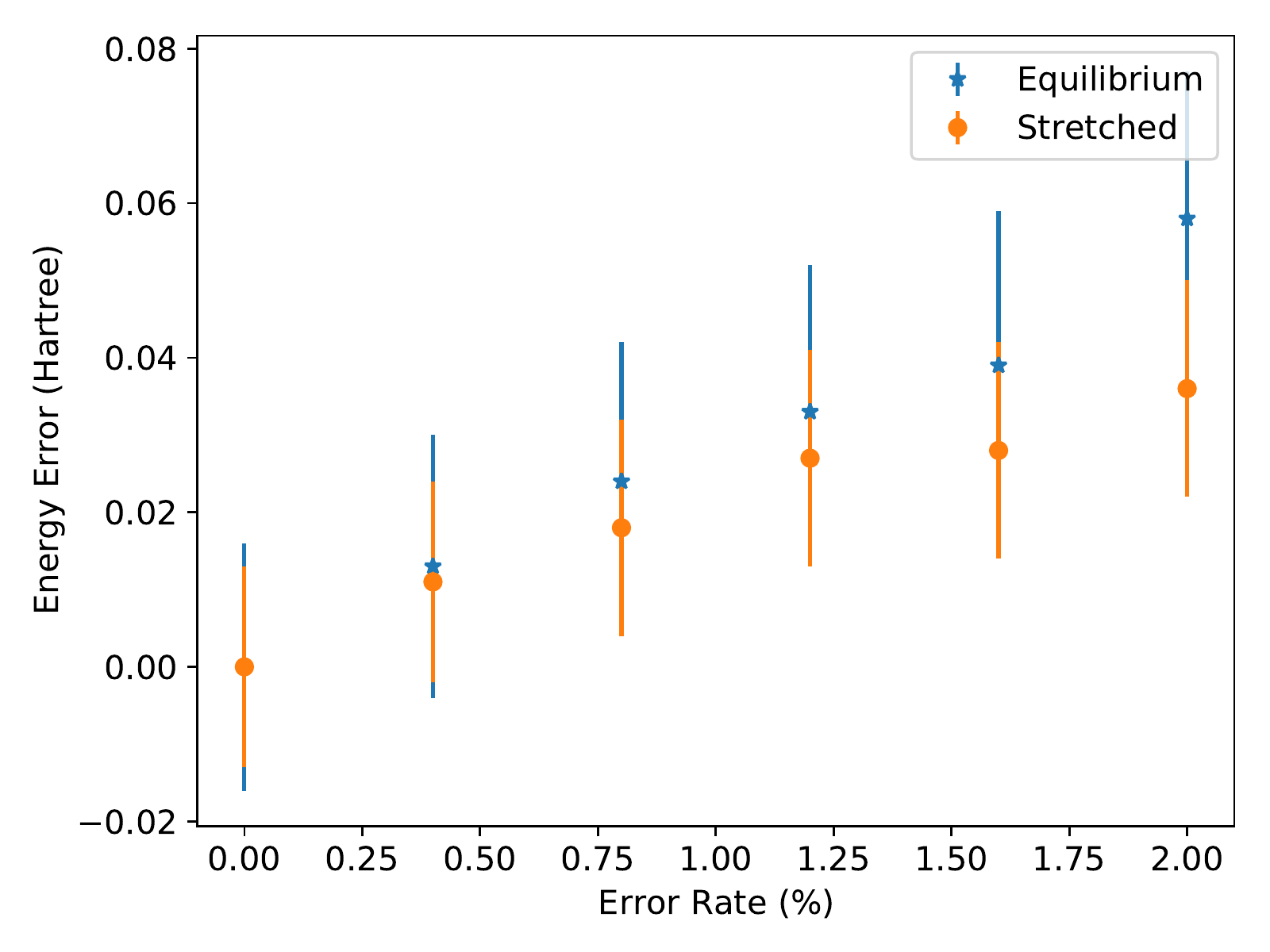}
\caption{
  VQE energy error as a function of error rate in the incoherent error model.
        }
\label{fig:incoherent_error}
\end{figure}

We then use this error channel for $R_{xx}$ gate in the $CX$ decomposition to simulate the same
systems as in the coherent noise case, and the results are shown in Figure \ref{fig:incoherent_error}.

As we can see from both figures, when the error rate is small enough, the energy errors for the 
two geometry points are indistinguishable, which jusifies the constant shift we used in the main text
to correct the VQE experimental energy.

\clearpage
\bibliography{Journal_Short_Name, main}